\documentclass{ws-ijmpa}
\usepackage{times}

\newcommand{\E}{E_{0}}
\newcommand{\Eth}{E_{\mathrm{th}}}
\newcommand{\rhom}{\rho_{\mathrm{\mu}}}
\newcommand{\rhos}{\rho_{\mathrm{s}}}
\newcommand{\sigmean}{\sigma_{\mu}^{\mathrm{mean}}}
\newcommand{\sigphys}{\sigma_{\mu}^{\mathrm{phys}}}
\newcommand{\sigmeth}{\sigma_{\mu}^{\mathrm{meth}}}

\begin{document}

\markboth{S.~P.~Knurenko et al.}
         {Muons with $\Eth \ge 1$~GeV \& Mass Composition in the Energy
          Range $10^{18}-10^{20}$~eV}
\catchline{}{}{}{}{}

\title{MUONS WITH $\Eth\ge1$~GEV AND MASS COMPOSITION IN THE ENERGY RANGE
       $10^{18}-10^{20}$~EV OBSERVED BY YAKUTSK EAS ARRAY}

\author{S.~P.~KNURENKO$^*$, V.~A.~KOLOSOV, I.~T.~MAKAROV, I.~YE.~SLEPTSOV, \\
        V.~R.~SLEPTSOVA and G.~G.~STRUCHKOV}
\address{Yu.~G.~Shafer Institute of Cosmophysical Research and Aeronomy
         Lenin Avenue 31, Yakutsk 677891, Russia \\
         ${}^*$s.p.knurenko@ikfia.ysn.ru
       }

\maketitle

\pub{Received (22 October 2004)}{Revised (4 July 2005)}

\begin{abstract}
The ratio of the muon flux density to charged particle flux density at distances 
of $300$ and $600$~m from the shower axis ($\rhom(300)/\rhos(300)$ and 
$\rhom(600)/\rhos(600)$) is measured. In addition, the energy dependence of 
$\rhom(1000)$ is analysed for showers with energies above $10^{18}$~eV. A 
comparison between the experimental data and calculations performed with the 
QGSJET model is given for the cases of primary proton, iron nucleus and gamma-
ray. We conclude that the showers with $\E\ge3\times10^{18}$~eV can be formed by 
light nuclei with a pronounced fraction of protons and helium nuclei. It is not 
excluded however that a small part of showers with energies above $10^{19}$~eV 
could be initiated by primary gamma-rays.

\end{abstract}

\smallskip
\smallskip

The Yakutsk Complex EAS array is able to record muons with the threshold
energy
\[
\Eth=1\,\mathrm{GeV}\times\sec(\theta).
\]
For this purpose the array is equipped with $6$ muon detectors with the full 
area of about $300$~m$^2$. The detectors are located at the distances
$R\sim100$, $300$, $500$ and $1000$~m from the array center.\cite{bib1}
The muon detectors are interrogated upon the availability of the array
``master'', i.e.\ the coincidence of responses from the scintillation detectors
of three observation stations located at the top of $500$~m arm equilateral
triangles. 
After almost $20$~years of continuous work of the array, a high statistics on 
the showers with energies above $10^{17}$~eV has been accumulated.

For the present study, the Yakutsk array data accumulated during the period from 
1994 to 2004 (i.e.\ after the array modernization\cite{bib1}) was taken into 
consideration. Thus the accuracy of the EAS axis determination has been 
increased twice. This considerably improved  the quality of the analyzed data. 
The showers with $\E\ge10^{18}$~eV and zenith angles $\theta<60^{\circ}$ have 
been chosen for the analysis.
\begin{figure}[t!]
\centering
\includegraphics[width=0.9\linewidth]%
 {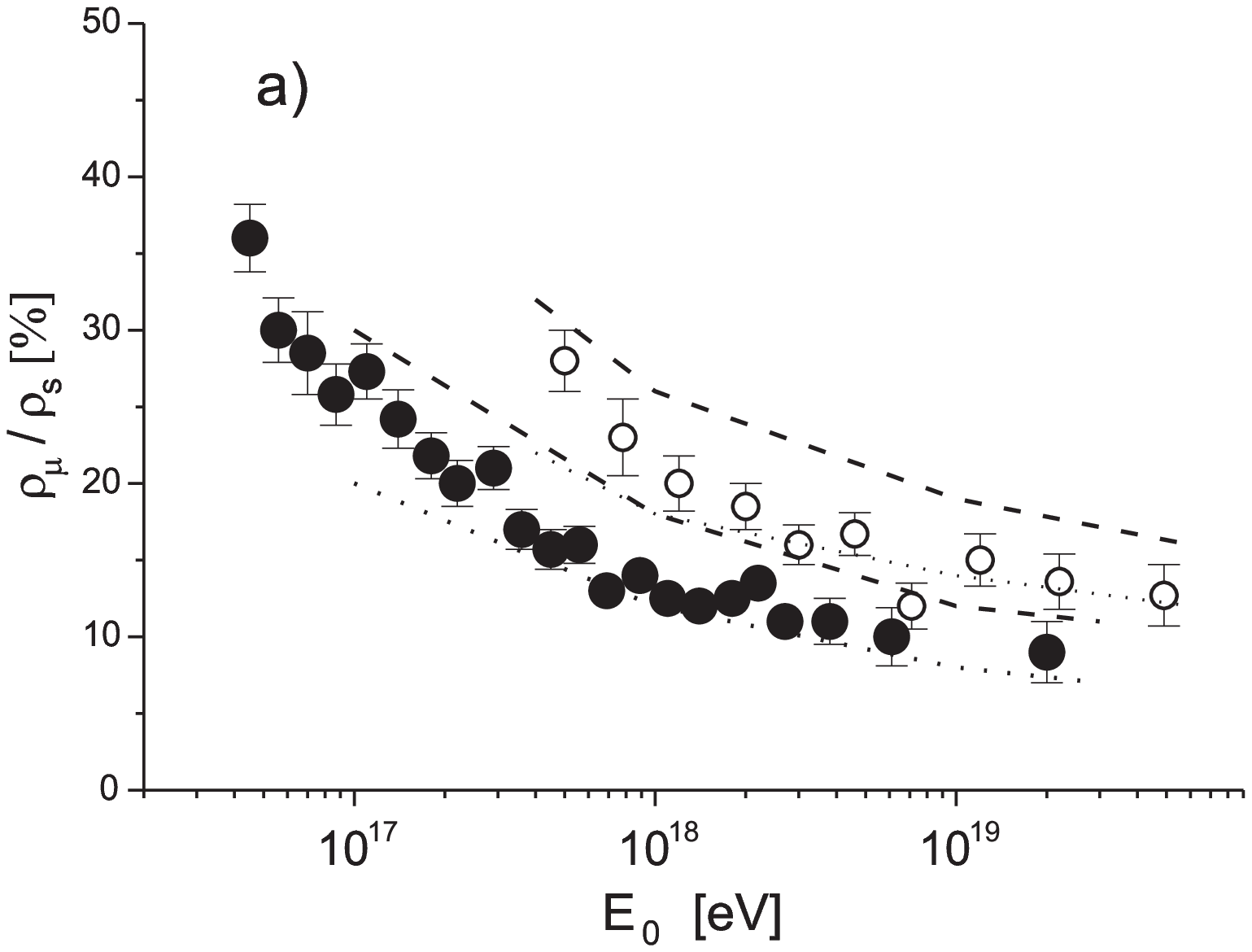} 
\includegraphics[width=0.9\linewidth]%
 {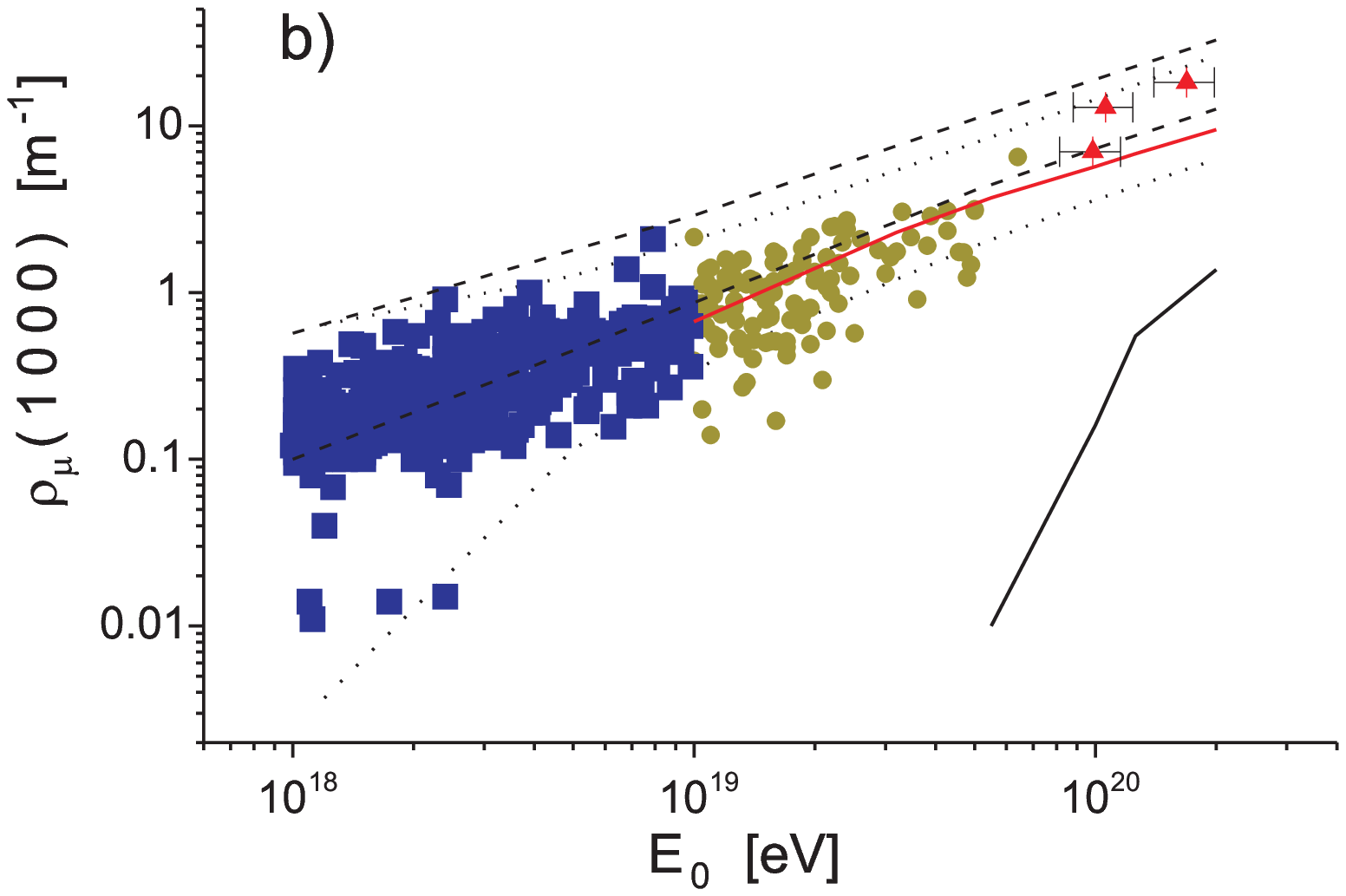}
\vspace*{8pt}
\caption{{\bf a)} Fraction of muons with energies above $1$~GeV~(\%):
         $\rhom(300)/\rhos(300)$ (\textbullet) and $\rhom(600)/\rhos(600)$ ($\circ$). 
         {\bf b)} $\rhom(1000)$  as a function of $\E$ for the observed events at
         energies within the ranges $10^{18}-10^{19}$~eV (squares),
         $10^{19}-10^{20}$~eV (filled circles) and $10^{20}-10^{21}$~eV (triangles).
         The expected $\pm1\sigma$ limits for the distributions are indicated
         (in both panels) by different curves for the proton, iron and gamma-ray
         primaries, as is explained in the text.
         }
\end{figure}
It was required for the shower axis to be within the boundaries of the array and 
that at least three muon detectors (one of which is at a distance of $1000$~m
from the axis) response to the shower. The muon flux density, $\rhom(1000)$,
is then determined as the median between the densities adjusted to
$\langle{R}\rangle=1000$~m in accordance to the mean muon lateral distribution
function. For this purpose, besides the detectors responded in the shower, the
data from the threshold detectors were also included into the analysis.
The muon flux densities at the distances $300$ and $600$~m from the shower axis
were calculated similar way. The results of our study are shown in Figs.~1\,a
and b.

Calculations suggest\cite{bib2,bib3} that the function $\rhom(R)$ slowly depends
on the zenith angle in the angular interval $\Delta\theta=(0^{\circ}-60^{\circ})$.
Hence, within this interval it will be is a reasonable approximation to apply
the mean angle $\bar\theta=39^{\circ}$ for a comparison of calculations and
experimental data. In order to take into account 
both physical and methodical fluctuations in the measurements of $\rhom(1000)$ 
(including the implementation errors and the errors in determination of the 
shower axis location) we considered the ``judicial'' interval in one 
$\sigmean=\sigphys +\sigmeth$.

In Figs.~1\,a and 1\,b the results of calculations are shown by dotted lines for 
the case of primary proton and by dashed lines for iron nucleus. Solid lines 
indicate the upper and lower limits for the EAS with $\E\ge10^{19}$~eV initiated 
by a primary $\gamma$-quantum. In this case we adopted the results of 
calculations from Ref.~\refcite{bib4}. The dots and circles in Fig.~1\,a are for 
the experimental data on $\rhom(300)$ and $\rhom(600)$, respectively. In 
Fig.~1\,b, the showers in the energy range of $10^{18}-10^{19}$~eV are shown by 
squares, the showers with $\E\ge10^{19}$~eV are shown by dots, and the showers 
of maximum energy are shown by triangles.

Our comparison of experimental data presented in Fig.~1\,a with the calculations 
carried out for the case of the primary proton and iron nucleus and based upon 
the QGSJET model, confirms the hypothesis that a considerable fraction of 
ultimate energy EAS is being formed by protons. The proton fraction of the 
primary cosmic rays decreases below the energy of about $10^{18}$~eV. It can be 
seen from Fig.~1\,b that the basic amount of points relevant to the showers 
having energies larger than $10^{19}$~eV falls into the ``proton interval''; 
$23$ points fall into the zone of superposition of primary protons and iron 
nuclei; and only $19$ showers among $116$ fall into the zone near the upper 
boundary evaluated under assumption that the primary particle is a
$\gamma$-quantum. Our analysis indicates that, even if the fluctuations were 
taken into account, a substantial probability remains that the showers of such 
energies are generated by neutral particles, in particular, by primary $\gamma$-
quanta. The analysis of arrival directions of showers with $\E \ge 10^{19}$~eV 
is justified for searching the sources of highest energy cosmic rays. The 
accuracy of determination of the arrival angles for these showers is expected to 
be not worse than ($0.5^{\circ}-1.5^{\circ}$).

\section*{Acknowledgements}

This work has been financially supported by RFBR, grant \#02--02--16380, grant
\#03--02--17160 and grant INTAS \#03--51--5112.

\end{document}